\documentclass[prb,aps,epsf,twocolumn,showpacs,superscriptaddress]{revtex4}
\usepackage{graphicx}
\usepackage{latexsym}
\usepackage{amsmath}

\newcommand{\barrop}{\bar{\bar{\rho}}^p({\bf q})}
\newcommand{\barropy}{\bar{\bar{\rho}}^p}
\newcommand{\barchi}{\bar{\bar{ \chi}}}

\newcommand{\beq}{\begin{equation}}
\newcommand{\beqr}{\begin{eqnarray}}
\newcommand{\eeqr}{\end{eqnarray}}
\newcommand{\eeq}{\end{equation}}

\newcommand{\etab}{\mbox{\boldmath $\eta $}}

\def\bp{{\mathbf p}}

\def\bq{{\mathbf q}}
\def\bp{{\mathbf p}}
\def\br{{\mathbf r}}

\def\bq{{\mathbf q}}

\def\bPi{{\mathbf \Pi}}

\def\bR{{\mathbf R}}
\def\bz{{\mathbf z}}

\def\half{{1\over2}}

\def\eqa{\begin{eqnarray}}
\def\eea{\end{eqnarray}}
\begin{document}
\title{The last word in strong correlations\\
}

\author{R. Shankar}
\affiliation{Department of Physics, Yale University, New Haven CT 06520}

\date{\today}

\begin{abstract}

In the Fractional Quantum Hall Effect (FQHE), in the noninteracting limit,  only a fraction $\nu $ of  the Lowest Landau Level (LLL) is occupied,  producing  a huge degeneracy. Interactions  lift this degeneracy and mix in higher  LL's. In the limit in which we ignore all but the LLL (i.e., let the inverse electron mass ${ 1 \over m}\to \infty$) , the kinetic energy is an irrelevant constant and
 the ratio of potential to kinetic energy is essentially infinite, making this the most strongly correlated problem imaginable. I give a telegraphic review of  the Hamiltonian Theory of the FQHE developed with Ganpathy Murthy  that  deals with this problem with some success. A nodding acquaintance with FQHE physics is presumed.\\
 {\em  To appear in Ann. Phys. {\bf 523}, 751, (2011), Dedicated to Dieter Vollhardt . }
\end{abstract}
\maketitle
\section{\label{intro}\ \ Introduction}

The first  breakthrough in the FQHE came from the approach
 pioneered by Laughlin \cite{laugh} and extended primarily  by Jain \cite{jain}
 and consists of writing down inspired trial wavefunctions.

The  hamiltonian approach \cite{rsgmrmp} is a complimentary one that begins with the microscopic hamiltonian for
interacting electrons and tries to obtain a satisfactory description
of the underlying physics through a sequence of transformations
and approximations.
It gives  a concrete operator
realization of many heuristic pictures that have been espoused and
makes precise under what conditions and in what sense these
pictures are valid. It allows   one  to compute to  reasonable accuracy
($10-20\%$ )  a large number of quantities such as gaps, relaxation rates, polarizations etc., at zero and nonzero
temperatures, at equal and unequal times, and even permits a crude model of disorder \cite{jim,usjim}.

The hamiltonian approach  differs from  the Chern-Simons (CS) approach  which has similar objectives. There one makes a
 singular gauge transformation on the electronic wavefunction (in the operator approach) \cite{lm} or couples electrons to a Chern-Simons gauge field (in the path integral approach) \cite{zhk,fradl},
leading in either case  to a composite particle which is the union of an electron
and some number of point flux tubes.

For the Laughlin fractions
\beq
\nu=\frac{1}{2s+1}
\eeq
where there are $2s+1$ external flux quanta per electron, one can either attach  $2s+1$ flux tubes to each electron in opposition to the applied field, and turn it into a boson in zero (mean) field that becomes a superfluid  \cite{zhk} or add $2s$ opposing flux quanta to turn it into a composite fermion that sees a net of one flux quantum and fills the LLL of CF's.

For the more general Jain fractions
\beq
\nu = {p \over 2ps +1},
\eeq
where there are $2s+{1 \over p} $ external flux quanta per electron, the only viable option is to   attach $2s$ quanta to each electron, producing  CF's that see ${1 \over p}$ flux quanta each and fill exactly $p$ LL's. This leads to the wavefunction  l

\beq \Psi= {\cal P} \prod_{j< i}(z_i-z_j)^{2s} \times \chi_p (z, \bar{z}) \exp (- \sum_i
|z_i|^2/4l^2)\label{jain}. \eeq
Here  $l=\sqrt{\hbar/eB}$ is   the electron's magnetic length.
The Jastrow factor up front comes from the flux attachment, $\chi_p (z, \bar{z})$ is the wavefunction for  $p$ filled CF LL's,  and the exponential factor is ubiquitous.
The operator ${\cal P}$ projects away all the $\bar{z}$ dependence of $\chi_p$ for $p>1$ so that $\Psi$ resides in the electronic LLL. In fact another projection has already been done: the flux- attaching gauge transformation  produces  only  the {\em phase}  of the Jastrow factor and the analytic zeros of the final Jastrow factor come only upon  projecting away the $|z_i-z_j|^{-2s}$ dependence of the phase.

The  $2s$-fold vortex at the location of every electron causes a charge deficit of $-2pse/(2ps+1)$,  as can be shown by Laughlin's plasma analogy or flux threading argument. Together with the charge $e$ of the electron, this implies a  screened quasiparticle of  charge $e^*=e/(2ps+1)$. It has no memory of the electron mass $m$, is sustained by just the interactions  and lives entirely in the LLL. This is the physical CF that we want to access and  describe in our appraoch.

A parameter that plays a central role in this article is
\beq
c^2= {2ps \over 2ps+1}\le 1.
\eeq

Henceforth we will focus on $s=1$ so that just two flux quanta are attached  to each electron in the CS approach and a double vortex appears at each electron's location in the wave function.

 The composite fermions of  CS theory   have a non-degenerate ground state at mean-field level, which is their main allure. This state is usually gapped.   The concept is however very effective even for the gapless case $\nu ={1 \over 2}$ ($c=1,  p=\infty$),  where many
phenomena pertaining to an over-damped mode, coupling to surface
acoustic waves, and the compressibility are  successfully described \cite{hlr}.

On the other hand, these CS fermions  do not exhibit in any
transparent way the quasiparticle properties (such as charge $e^*$ or
effective mass $m^*$ ) deduced from trial wavefunctions, do not reside in the LLL and indeed have a singular limit as  the electron mass $m\to 0$.

The  Hamiltonian Theory  which Murthy and I developed
over the years \cite{rsgmrmp} works in the LLL all along and  addresses some of these issues. We asked how one is to incorporate into the theory what the interpretation of the excellent wavefunctions tells us, namely that "an electron is bound to two vortices."
But what does that even mean?
 Vortices are not elementary particles with their own degrees of freedom or dynamics, they are zeros in a wavefunction for electrons! In our earlier work we  dealt with this  by introducing  extra plasmonic degrees of freedom {\em $\acute{a}$ la } Bohm-Pines \cite{bp} in an enlarged Hilbert space with some constraints.   We wrote down a product wavefunction (for electrons and plasmons),  which  upon projection to the (constrained) physical sector   yielded  the Jain wave function including the complete analytic Jastrow factor {\em with its multiple    zeros}  and  not just its phase. This approach evolved  and got refined over time. Here I fast-forward to the final version which is an efficient   starting point for many calculations, directing the reader eager  for details to Ref. \cite{rsgmrmp}.

\section{The Hamiltonian approach}

The
 primordial hamiltonian in { terms} of electronic variables
(which carry the subscript $e$ to distinguish them from other
coordinates to be introduced shortly) is: \beq H = \sum_j
{\etab_{ei}^{2}\over 2ml^4}+{1\over 2}\sum_{i,j,\bq }v(q)e^{ i\bq
\cdot (\br_{ei} - \br_{ej})}\equiv H_0 +V \label{ham}\eeq
where  $\etab_{ei}$ is the i-th electron's cyclotron coordinate that lets it move up and down Landau levels and  $\br_{ei}$ is its coordinate in the plane.

Projecting  to
the LLL  one drops the first (kinetic energy ) term and makes
the replacement
\beq
e^{i \bq \cdot {\br_e}} = e^{i \bq \cdot ({\etab_e+\bR_e})}\to \langle  e^{i \bq \cdot {\etab_e}}\rangle_{\mbox{\small LLL}} e^{i \bq \cdot {\bR_e}} =e^{-q^2l^2/4}\cdot e^{i \bq \cdot {\bR_e}}
\eeq
where $\bR_e$ is the electronic guiding center coordinate. Thus the projected Hamiltonian is
\beq
\bar{\bar{H}}_{} =\half \sum_{i,j,\bq }
v(q)\ e^{-q^2l^2/2} \ e^{i\bq \cdot (\bR_{ei} - \bR_{ej})}
\eeq

(While we limit  ourselves to the LLL in this article,  one can  retain the   $\etab_e$ in Eqn. (\ref{ham}) to  study, for example,  LL mixing as we did in Ref.  \cite{llmix}.)

Although we do not have a kinetic energy term, the problem is hard because the components of $\bR_e$ do not commute
\beq
\left[ R_{ex}, R_{ey}\right]=-il^2.
\eeq

Thus $R_{ex}, R_{ey}$ are conjugate variables. Now a full fledged fermion in $d=2$ will have two coordinates and two momenta, i.e., two conjugate pairs. The LLL projected electron is like half a fermion and this is what makes the analysis difficult. So we introduce another conjugate pair of "vortex" guiding center coordinates which we will define by their commutation relations:
\beq
\left[ R_{vx}, R_{vy}\right]={il^2\over c^2} \ \ \ \ \mbox{where    $c^2={2p\over 2p+1}$.}
\eeq
Thus the vortex describes a particle whose charge  $-{2p \over 2p+1}$ in electronic units is exactly that of the vortices in the Jastrow factor. It too is just half a particle like the projected electron.

We want these extraneous coordinates to commute with everything electronic i.e.,
\beq
\left[ \bR_{e}, \bR_{v}\right]=0.
\eeq

{\em Now the point is that we can accommodate both $\bR_e$ and $\bR_v$ and their algebra very neatly  into the   Hilbert space of a regular two-dimensional  fermion, which is going to be  our  composite fermion.}    This fermion is bathed in the reduced field seen by a  $e^*$ object.  From its position vector $\br$ and kinetic momentum  ${\bf \Pi } = {\bf p} -e {\bf A}
$ we can construct  its guiding center and cyclotron coordinates (which carry no subscripts like $e$ or $v$)  that  obey

\begin{eqnarray}
\left[ \eta_x , \eta_y \right]&=& il^{*2} = {il^2\over 1-c^2}\\ \left[
R_x , R_y \right]&=& -il^{*2}.
\end{eqnarray}

In terms of these two conjugate pairs
$\bR_e$ and $\bR_v$  can be represented  as follows:

\begin{eqnarray}
\bR_e &=& \bR + \etab c \\ \bR_v &=& \bR +\etab /c.
\end{eqnarray}

An equivalent representation in terms of $\br$ and $\bPi$, the CF coordinate and velocity
operators,  is \begin{eqnarray}
 \bR_e &=&    {\bf r} -{l^2\over (1+c)}\hat{\bf z}\times {\bf \Pi
},   \label{straddle1} \\
 \bR_v &=&{\bf r} +{l^2\over c(1+c)}\hat{\bf z}\times {\bf \Pi
}\label{straddle2}.
\end{eqnarray}

The inverse transformation is illuminating:
\begin{eqnarray}
\bR &=& {\bR_e -c^2 \bR_v \over 1-c^2} \\ \etab &=& {c \over
1-c^2} (\bR_v-\bR_e)
\end{eqnarray}
The first equation could have been guessed: it says  that the  CF guiding center is
the weighted sum of its charged parts.   The second equation can be found by
demanding that $\etab$ be linear in $\bR_e$ and $\bR_v$, commute
with $\bR$,  and have an overall scale that produces the right
commutator.

 Consider Eqns. (\ref{straddle1},\ref{straddle2})   when $\nu={1 \over 2}$ or $c=1$, and ${\bf \Pi } = \bp$ (the CF sees no field).  We see that  $\bR_e$ and $\bR_v$ are located on either side of $\br$ separated by   $\hat{\bf z}\times\bp l^2$. This is the operator realization of Read's  dipole picture \cite{dipole}.

 Ignoring the zero point energy, here is where we stand in the LLL
 sector:
\begin{eqnarray} \bar{\bar{H}}_{} &=&\half \sum_{i,j,\bq }
v(q)\ e^{-q^2l^2/2} \ e^{i\bq \cdot (\bR_{ei} - \bR_{ej})}\\ &=&
\half \sum_{i,j,\bq } v(q)\ e^{-q^2l^2/2} \exp \left({i\bq \cdot
\left[ (\bR_{i} - \bR_{j})+c(\etab_i-\etab_j)\right]}\right)
\end{eqnarray}

While it is true that we have managed to get rid of the electron mass $m$ and
isolate the LLL cleanly, the reader may ask what we have gained,
since {\em algebraically} the problem is the same as in electronic
coordinates. {\it The answer is that now there is a natural
nondegenerate HF ground state in the enlarged CF  space}. This is
because the HF hamiltonian is now  written in terms of CF
operators $\bR$ and $\etab$ and  the particle density is just
right to fill  the lowest  $p$  CF-LL's. This  key step  opens up all the usual
approximation schemes.

From $\bR_e$ and $\bR_v$ we can form the corresponding electron and vortex densities:

\beq \bar{\bar{\rho}}({\bf q}) = \sum_j e^{-i\bq
\cdot \bR_e}\eeq
and
\beq \barchi  = \sum_j e^{-i\bq
\cdot \bR_v}. \eeq
These  obey
\beq \left[
\bar{\bar{\rho}}({\bf q}) , \bar{\bar{\rho}}({\bf q'}) \right] =2i
\sin \left[ { ({\bf q\times q'})\ l^2 \over 2}\right]
\bar{\bar{\rho} }({\bf q+q'}) \eeq which was thoroughly exploited
in  \cite{gmp}
and
\begin{eqnarray}
\left[ \bar{\bar{\chi}}(\bq) \ ,  \bar{\bar{\chi}}(\bq') \right]
&=& -2i \sin \left[ {l_{}^{2} ({\bf \bq\times \bq'}) \over
2c^2}\right] \bar{\bar{\chi}} (\bq +\bq').
 \end{eqnarray}

The electron and vortex densities commute since $\bR_e$ and $\bR_v$ do.

The mathematical problem we face is then summarized by the following:
\begin{eqnarray}
\bar{\bar{H}}  &=&\half \sum_{\bq } v(q)\ e^{-q^2l^2/2} \bar{\bar{\rho}}({\bf q})
\bar{\bar{\rho}}(-\bq)\label{cons1}\\
\left[ \bar{\bar{H}}\ , \barchi \right] &=&0\label{cons2}\\
\barchi &\simeq & 0
\end{eqnarray}
where the last equation need some explanation. Since $\bR_v$ and $\barchi$ do not appear in $\bar{\bar{H}}$, $\barchi$ does  not have any dynamics, just like the longitudinal part of the vector potential in a gauge theory where the hamiltonian is gauge invariant.
We shall demand that
 $\barchi \simeq 0$ which means $\barchi$ will vanish within correlation functions. (Since $\barchi$ commutes with $\bar{\bar{H}}$, this  is a first class constraint preserved  by the equations of motion. )

 Our equations above are good for all Jain fractions. Similar equation for the case where the vortex and CF had equal and opposite charges  were written down independently by Pasquier and Haldane \cite{ph} and extended and exploited by Read \cite{read2} to address the  $\nu =\half $ problem.

 \section{Putting the Hamiltonian Theory to work}

   The rationale for working with the CF was to get a unique ground state for the HF approximation.

There are at least two good reasons to expect that the naive HF
result will require fairly strong  corrections. First,  if we
compute the matrix element of the projected electron density
between any two HF states, the answer will be  linear in $q$,
whereas in the exact theory, we know that within the LLL it must go as
$q^2$ as per Kohn's theorem \cite{kohn}. Secondly,  as $ql \to
0$, the projected electronic density  has unit contribution from each CF
while we would like it to be $e^*/e={ 1\over 2p+1}=1-c^2$. Evidently the HF result will
receive strong corrections that will renormalize these quantities
till they are in line with these expectations. These renormalization will occur once we pay attention to the constraint $\barchi \simeq 0$.

Now  Baym and Kadanoff \cite{bk}) have a   procedure for
improving the HF state with additional diagrammatic corrections (ladder sums) to enforce conservation laws. (The non-conservation comes from using Hartree- Fock self-energies for propagators while using bare vertices in the one-loop response functions, in violation of Ward-identities.)

For $\nu ={1 \over 2}$ Read \cite{read2}
 showed that this procedure restores Kohn's theorem, exhibits the overdamped mode, reveals
a dipolar structure for density-density correlations and yield a compressible state.  Murthy \cite{gmcon} has used it to calculate density-density correlations in gapped fractions. We are currently using it to perform a  comparative study of the fractions $\nu ={1 \over 2}$ and ${5 \over 2}$. In general this route must be followed whenever the constraint (or gauge invariance under the transformations generated by $\barchi$ ) is important.

\subsection{ Preferred charge: a short cut to the constraint}
We found that in many problems where there is  a large enough   gap,  temperature or disorder there is a short-cut to implementing the constraint in the infrared limit. We discuss this in some detail, for unlike the Baym-Kadanoff route, this one  is peculiar the FQHE and  we do not fully understand why it works
or how we could give a better interpretation for it.

 Suppose, in the hamiltonian and elsewhere,   we  replace $\bar{\bar{\rho}}({\bf q})$ by the {\em
preferred combination} \beq  \barrop = \bar{\bar{\rho}}({\bf q}) -c^2 \bar{\bar{\chi
}} .\eeq
In an exact calculation it  make  no difference to the computation of anything physical whether the  coefficient in front of  $\barchi$  is zero or $-c^2$ or anything else   since $\bar{\bar{\chi }}$ is
essentially zero.

On the other hand in the HF approximation (which does not respect $\barchi \simeq 0$) it certainly matters what coefficient we place  in front of  $\barchi$. The preferred combination  $\barrop$ stands out as   the  sum of the electronic and vortex charge densities. But the reason we are forced to use is that it helps us avoid  violating  Kohn's Theorem within simple HF.

 Consider its expansion  in powers of $ql$: \beq \bar{\bar{\rho}}^p
= \sum_j e^{-i{\bf q \cdot r}_j}\!\! \left(\! {1 \over 2p+1} \!
-     {i l_{}^{2} } \bq\times {\bf \Pi}_{j}  \! +{0} \cdot \left(
\bq \times {\bf \Pi}_{j} \right)^2 \! + \cdots \right)
.\label{rostar1} \eeq

\begin{itemize}
\item
The   transition matrix elements are now of order $q^2$ between HF
states because  coefficient of $\bq$ is proportional to the CF
guiding center coordinate $ \br - l^{*2}\hat{\bz}\times \bPi$ with
no admixture of the CF cyclotron coordinate. This is more
transparent if we use $\bR$ and $\etab$ to write \begin{eqnarray}
\barrop &=& (1 -i \bq \cdot (\bR + c \etab) + ..)-c^2 (1 -i \bq
\cdot (\bR +  \etab /c +...)\\ &=& (1-c^2)(1 - i\bq \cdot \bR +
{\cal O} \ (q^2).\end{eqnarray}
The choice of  ($-c^2$)  as the coefficient of $\barchi $ in $\barrop$, uniquely determined by compliance with Kohn's Theorem,  is also the one that leads to two important  collateral benefits:
\item The electronic charge
 density
associated with $\barrop$ is  $1-c^2 =e^*/e$.
\item We see from Eqn. (\ref{rostar1}) that when $\nu =\half $, the preferred density
couples to an external electric field like a dipole of size $d^*=l^2
\hat{\bz}\times \bp $ giving a precise operator expression of  Read's picture \cite{dipole}. \end{itemize}

The hamiltonian ${\bar{\bar H}}(\barropy ) $  is {\em weakly
gauge invariant}, that is \beq \left[ {\bar{\bar H}}(\bar{\bar{\rho}}^p) \ ,
\barchi \right] \simeq 0 \eeq where the $\simeq 0 $ symbol means
that it vanishes in the subspace obeying $\barchi =0$. Thus
neither ${\bar{\bar H}} (\bar{\bar{\rho}}^p)$ nor $\bar{\bar{\rho}}^p$ will mix
physical and unphysical states.

The      significance of ${\bar{\bar H}}(\bar{\bar{\rho}})$   is the
following. If the constraint      $\bar{\bar{\chi}} =0$ is imposed
{\em exactly},       there are many equivalent hamiltonians
depending on how $\bar{\bar{\chi}}$ is insinuated into it.
However, in the HF {\em approximation}, these are not equivalent
and ${\bar{\bar H}}(\bar{\bar{\rho}}^p)$ best approximates, between HF states
and at long wavelengths, the true hamiltonian between true
eigenstates.  In contrast      to a variational calculation where
one searches among trial states      for an optimal one, here the HF
states are the same for a      class of hamiltonians (where
$\bar{\bar{\chi}}$ is      introduced into ${\bar{\bar H}}$      in any
rotationally invariant form),  and we seek the best hamiltonian, which happens to be
${\bar{\bar H}}(\bar{\bar{\rho}}^p)$ since encodes the fact that every electron is
accompanied by a correlation hole of some sort, which leads to the correct  $e^*$, $d^*$, and obeys the all important   Kohn's theorem
 ($q^2$ matrix elements for the  density projected to the LLL.)

 The preferred charge $\barrop$ and preferred hamiltonian ${\bar{\bar H}}(\bar{\bar{\rho}}^p)$ have been used to compute  gaps, finite temperature response functions (polarization, NMR rates) and even the effect of disorder. The results are in reasonable agreement ( $10-20 \%$ )    with computer simulations and real data \cite{rsgmrmp}.

Note that when we use the preferred charge and hamiltonian we
make no further reference to constraints, and simply carry out the
Hartree-Fock approximation. This is based on the expectation that even
if we found some way to include the effect of constraints, it will
make no difference in the small $ql$ region because the
leading renormalization of $e$ to $e^*$ and suppression of ${\cal} \
q$ matrix elements down to ${\cal} \ q^2$ that are achieved by the
conserving approximation  are
built in here. Of course errors at larger $q$ will corrupt the actual numbers, say for gaps.

The shortcut however fails in one important regard. For the gapless  $\nu =\half$ state at $T=0$, since
$\barrop$ starts out linearly in $\bq$, the CF couples like a dipole to the external potential, leading to a  compressibility that vanishes as $q\to 0$. The only way to restore compressibility is to have some very low energy collective excitations that overcome the factors of $q$ in the matrix elements. This was first   pointed out to us by Halperin and Stern in \cite{stern0} who used a toy model to make their point that respecting gauge invariance (or the constraint) is crucial. They  went on to give a  detailed analysis of the realistic model with additional  coworkers \cite{stern}.
 Subsequently Read \cite{read2} did the  ladder sum on top of HF and obtained the overdamped mode, finite compressibility and dipolar coupling.

 The reader will recall that any {\em simple} picture of
quasiparticles, whether it be in Landau's Fermi liquid theory, or
in BCS theory, is best captured by approximate and not exact
descriptions. The quasiparticles are all caricatures of some exact
reality and therein lies their utility. Similarly the CF in our
extended formalism appears only in the HF approximation to
${\bar{\bar H}}( \barropy) $. Recall that we brought in the coordinate $\bR_v$ to
become the electron's partner in forming the CF.  However $\bR_v$
was cyclic in the exact hamiltonian ${\bar{\bar H}}$. {\em Thus the exact
dynamics never demanded that $\bR_v$ be bound to $\bR_e$ or even
be anywhere near $\bR_e$. } However, in the HF approximation,
since we wanted the right charge and transition matrix elements of
the density operator (Kohn's theorem) to be manifest, we needed to
replace $\bar{\bar{\rho}}$ by $\barropy$, and trade ${\bar{\bar H}}(\bar{\bar{\rho}})$ for ${\bar{\bar H}}(\barropy)$,
the preferred hamiltonian. In ${\bar{\bar H}}(\barropy )$, $\bR_v$ is coupled to
$\bR_e$. The HF approximation and this coupling go hand in hand.
The exact eigenfunctions of the original ${\bar{\bar H}}$ are factorized in
the analytic coordinates $z_e$ and $z_v$ and presumably reproduce
the electronic correlations of the FQHE states.  On the other
hand, in the HF approximation to ${\bar{\bar H}}(\barropy )$, the wavefunctions
(e.g., $p$-filled LL's) mix up $ z_e$ and $z_v$, and ${\bar{\bar H}}(\barropy  )$,
the preferred hamiltonian, dynamically couples $\bR_e$ and
$\bR_v$. The net result is that, at least at long wavelengths,
these two wrongs make it right and mimic what happens in the exact
solution.

Another advantage of ${\bar{\bar H}}(\barropy )$ is that it gives an approximate
formula for $m^*$ originating entirely from interactions. This is
best seen at  $\nu =\half$.  When we square $\bar{\bar{\rho}}^p$
(Eqn. (\ref{rostar1}), we get a double sum over particles whose
diagonal part is the one particle (free-field) term: \beq
H^{0}_{\nu ={1\over 2}}=2\sum_j \int {d^2q\over 4\pi^2} \sin^2
\left[{{\bf q \times k }_j l^2\over 2}\right] {v}(q)
e^{-q^2l^2/2}. \label{freeh} \eeq

This is not a hamiltonian of the form $k^2/2m^*$. However if the
potential is peaked at very small $q$, we can expand the sine and
read off an approximate $1/m^*$

\beq {1\over m^*}= \int {qdq d \theta \over 4\pi^2} \left[ (\sin^2
\theta)\ (ql)^2 \right] {v} (q)\ e^{-q^2l^2/2} \eeq which has its
origin in electron-electron interactions. However we can do more:
we have the full $H_0$ as well as the interactions. The point to
emphasize is that $H$ is not of the traditional form ($p^2/2m +V$
) and that there is no reason it had to be. This proves crucial in understanding the data from Ref. \ref{j}.

\section{Conclusions and Summary}

The trial wavefunctions tell us that the quasiparticle of the FQHE is an electron bound to vortices, that this entity which resides in the LLL of electrons has charge $e^*$, mass and dynamics generated entirely by the interactions and no memory of $m$.  Here we show one way to implement that concept within a Hamiltonian and  commuting constraints.  It consists of complementing $\bR_e$, the guiding center coordinate of the electron with $\bR_v$, the guiding center coordinate of an entity that has the same charge as the vortex. The two guiding centers fit nicely into the Hilbert space of the composite fermion which sees a field just right to fill $p$ LL's or the Fermi sea, paving the way for a HF calculation. But one finds that naive HF violates the constraint, violates Kohn's theorem and describes a particle with the electronic charge $e$ and not $e^*$.

We describe   two ways to fix it. One is the standard ploy of  using a conserving approximation in which particle-hole ladder graphs restore the Ward identity.  This is quite involved but necessary whenever physics at very small $\omega$ and $q$ is to be faithfully described and gauge invariance is crucial.

The second method, which seems to have no analog outside of the FQHE is to add to the  electronic density in the Hamiltonian a judicious  amount of the constraint to salvage Kohn's theorem within the naive HF calculation. This ends up producing the right quasiparticle charge and  dipole moment. It describes the data on gaps, relaxation rate and polarization rather well  at finite frequency, wave vector, temperature and even disorder.
 At this moment we do not have a deeper understanding of why it works as well as it does or if there is a better way to introduce it into the formalism. I offer it as a challenge to the readers, especially Dieter, to stimulate his aging brain.

\section{Acknowledgements}

I  thank the NSF for grant DMR- 0901903, Ganpathy Murthy for his collaboration and finally Dieter Vollhardt, for years of warm friendship, matchless hospitality  and constant enlightenment.


\begin{thebibliography}{8.}
\bibitem{laugh} R.B.Laughlin,  Phys. Rev. Lett. {\bf 50},
1395, (1983)
\bibitem{jain} J.K.Jain, Phys. Rev. Lett.  {\bf 63}, 199, (1989), J.K.Jain and R.K.Kamilla, in Chapter 1 of, {\it
``Composite Fermions''}, Olle Heinonen, Editor (World Scietific,
Teaneck, NJ, 1998). 
\bibitem{rsgmrmp} G. Murthy and R. Shankar, \rmp {\bf 75}, 1101 (2003).
\bibitem{jim} L. A. Tracy, J. P. Eisenstein, L. N. Pfeiffer, and K. W. West, Phys. Rev. Lett. {\bf 98}, 086801 (2007) describes the experiment at nonzero $T$ and disorder. \label{j} 
\bibitem{usjim} G. Murthy and R. Shankar,  Phys. Rev.
    {\bf B 76}, 075341, (2007) explains     the results of Ref. \ref{j} in the Hamiltonian picture. 
    \bibitem{lm} J. M. Leinaas and J. Myrheim, Nuovo Cimento {\bf 37B}, 1 (1977).
\bibitem{zhk}S.-C. Zhang, H. Hansson and S. A. Kivelson, Phys. Rev. Lett.
{\bf 62}, 82, (1989); D.-H. Lee and S.-C. Zhang, Phys. Rev. Lett.
{\bf 66}, 1220 (1991); S.-C. Zhang, Int. J. Mod. Phys., {\bf B6},
25 (1992).
\bibitem{fradl} A. Lopez and E. Fradkin, Phys. Rev. B {\bf  44}, 5246
(1991), {\em ibid} {\bf 47}, 7080, (1993), Phys. Rev. Lett.  {\bf
69}, 2126 (1992).
\bibitem{hlr} B. I. Halperin, P. A. Lee and N. Read, Phys. Rev. B {\bf
47}, 7312 (1993); See also, B. I. Halperin, in {\it Perspectives
in Quantum Hall Effects}, Das Sarma and Pinczuk, Editors (Wiley,
New York, 1997).
\bibitem{bp} D. Bohm and D. Pines, Phys. Rev. {\bf 92}, 609, (1953). 
 \bibitem{llmix} G. Murthy and R. Shankar, Ganpathy Murthy, Phys. Rev., {\bf  B 65},
245309, (2002).
\bibitem{dipole} N. Read, Semi. Sci. Tech. {\bf 9}, 1859 (1994);
Surf. Sci., {\bf 361/362}, 7 (1996).
\bibitem{gmp} S.M. Girvin, A.H. MacDonald and
P.Platzman, Phys. Rev. {\bf B33}, 2481, (1986)
\bibitem{ph} V. Pasquier and F.D.M. Haldane, Nucl. Phys., {\bf B156}, 719, (1998).
    \bibitem{read2} N. Read, \prb {\bf 58}, 16262 (1998).
\bibitem{kohn} W. Kohn, Phys. Rev {\bf 123}, 1242 (1961).
\bibitem{bk} G. Baym and L. P. Kadanoff, Phys. Rev. {\bf 124},
287 (1961); L. P. Kadanoff and G. Baym, {\it ``Quantum Statistical
Mechanics''}, Addison-Wesley, Reading, MA, 1989.
\bibitem{gmcon} G. Murthy, Phys. Rev. {\bf B 64},195310 , (2001).
\bibitem{stern0} This took place in the form of a comment by B. I. Halperin and A. Stern our response in
  Phys. Rev. Lett , {\bf 80   }, 5458,(1998).
\bibitem{stern} B.I. Halperin and A. Stern, Phys. Rev. Lett., {bf 80}, 5457, (1998), A. Stern, B.I. Halperin, F. vonOppen and S.H. Simon Phys. Rev. {\bf B59}, 12547, (1999).



\end{thebibliography}
\end{document}